\documentclass[conference]{IEEEtran}
\IEEEoverridecommandlockouts

\usepackage{cite}
\usepackage{amsmath,amssymb,amsfonts}
\usepackage{algorithmic}
\usepackage{graphicx}
\usepackage{textcomp}
\usepackage{xcolor}
\usepackage{amsmath,amsfonts}
\usepackage{array}

\usepackage[bookmarks=false]{hyperref} 
\usepackage{float}
\usepackage{textcomp}
\usepackage{stfloats}
\usepackage{url}
\usepackage{verbatim}
\usepackage{graphicx}
\usepackage{placeins} 
\usepackage{pifont}
\usepackage{booktabs}
\usepackage{multirow}
\usepackage[linesnumbered,ruled,vlined]{algorithm2e}
\RequirePackage{letltxmacro}
\LetLtxMacro{\LaTeXtextbf}{\textbf}
\LetLtxMacro{\textbf}{\LaTeXtextbf}
\usepackage{cite}
\usepackage{amsmath,amssymb,amsfonts}
\usepackage{graphicx}
\usepackage{textcomp}
\usepackage{color}
\usepackage{caption}
\usepackage{subcaption}
\usepackage{enumitem}  
\def\BibTeX{{\rm B\kern-.05em{\sc i\kern-.025em b}\kern-.08em
    T\kern-.1667em\lower.7ex\hbox{E}\kern-.125emX}}

\usepackage{tcolorbox}
\tcbuselibrary{breakable}

\begin{document}

\title{Towards Trustworthy Agentic IoEV: AI Agents for Explainable Cyberthreat Mitigation and State Analytics\\

}

\author{
\IEEEauthorblockN{Meryem Malak Dif, Mouhamed Amine Bouchiha, Abdelaziz Amara Korba, Yacine Ghamri-Doudane \\
\{meryem.dif, mouhamed.bouchiha, abdelaziz.amara\_korba, yacine.ghamri\}@univ-lr.fr
}
\IEEEauthorblockA{ L3i - La Rochelle University, La Rochelle, France}}

\maketitle

\begin{abstract}
The Internet of Electric Vehicles (IoEV) envisions a tightly coupled ecosystem of electric vehicles (EVs), charging infrastructure, and grid services, yet remains vulnerable to cyberattacks, unreliable battery-state predictions, and opaque decision processes that erode trust and performance. To address these challenges, we introduce a novel Agentic Artificial Intelligence (AAI) framework tailored for IoEV, where specialized agents collaborate to deliver autonomous threat mitigation, robust analytics, and interpretable decision support. Specifically, we design an AAI architecture comprising dedicated agents for cyber-threat detection and response at charging stations, real-time State of Charge (SoC) estimation, and State of Health (SoH) anomaly detection, all coordinated through a shared, explainable reasoning layer; develop interpretable threat-mitigation mechanisms that proactively identify and neutralize attacks on both physical charging points and learning components; propose resilient SoC and SoH models that leverage continuous and adversarial-aware learning to produce accurate, uncertainty-aware forecasts with human-readable explanations; and implement a three-agent pipeline, where each agent uses LLM-driven reasoning and dynamic tool invocation to interpret intent, contextualize tasks, and execute formal optimizations for user-centric assistance. Finally, we validate our framework through comprehensive experiments across diverse IoEV scenarios, demonstrating significant improvements in security and prediction accuracy. All datasets, models, and code will be released publicly.
\end{abstract}


\begin{tcolorbox}[breakable,boxrule=1pt,colframe=black,colback=white]
\scriptsize Paper accepted at 50th Annual IEEE Conference on Local Computer Networks (LCN'25) IEEE, 2025.
\end{tcolorbox}

\section{Introduction}
\label{sec:intro}
\quad \IEEEPARstart{T}{he} Internet of Electric Vehicles (IoEV) envisions a future in which electric vehicles (EVs) interact seamlessly with one another, with charging infrastructure, and with grid services through connected, data-driven systems. This vision is fueled by the ongoing modernization and globalization of the transportation sector, where EVs play a pivotal role in promoting greener and more sustainable mobility solutions.
As adoption grows, intelligent services such as State of Charge (SoC) estimation, State of Health (SoH) monitoring, and predictive maintenance are becoming increasingly essential for achieving energy efficiency, operational safety, and long-term reliability \cite{xie2023state, cao2025model}—drawing considerable attention from both industry and academia. In particular, accurate assessment of SoC and SoH is crucial, as these parameters directly influence battery utilization, driving range, and overall vehicle performance. SoC indicates the current charge level and guides decisions related to charging schedules and trip planning, while SoH reflects the battery’s degradation, aging, and performance, offering key insights for lifespan estimation and maintenance planning~\cite{mousaei2024advancing}. 

At the same time, the IoEV paradigm introduces broader system-level concerns, particularly the need for a secure and resilient charging infrastructure that can function reliably in dynamic and potentially adversarial environments. The very connectivity that enables intelligent services also expands the system’s attack surface, increasing its vulnerability to cyber threats and unpredictable operational disruptions~\cite{aljohani2024comprehensive}. As a result, IoEV systems must contend with multidimensional challenges that undermine their reliability, safety, and trust.
Among these, charging stations (CSs) are frequent targets for cyberattacks such as distributed denial-of-service (DDoS) or node compromise,  which can disrupt grid coordination, reduce service availability, and impair system-wide responsiveness~\cite{almadhor2025transfer}. In parallel, the accuracy of SoC and SoH predictions can deteriorate due to data heterogeneity, sensor failures, or adversarial inputs— leading to flawed assessments that compromise both safety and range estimation~\cite{mousaei2024advancing}. Furthermore, the increasing complexity of IoEV environments gives rise to human-in-the-loop concerns, where oversight is often reactive, opaque, or inadequately informed—partly due to the black-box nature of many learning-based models. These limitations underscore the need for autonomous and explainable solutions that can adapt to evolving threats, produce reliable predictions, and deliver transparent reasoning that can be understood by both machines and human operators.

To this end, we propose leveraging Agentic Artificial Intelligence (AAI), a class of AI systems capable of autonomous decision-making, contextual reasoning, and goal-directed adaptation. Agentic AI frameworks are well-suited for dynamic settings like IoEV, where agents must operate under uncertainty, interact with potentially compromised peers, and pursue long-term safety and efficiency objectives~\cite{dehrouyeh2024tinyml}. When augmented with explainability, Agentic AI not only enhances trust and accountability but also enables actionable insights for system operators and end users, bridging the gap between automation and human comprehension\cite{khowaja2025integration}.

Beyond battery diagnostics and infrastructure security, our framework introduces a user-centric dimension to IoEV intelligence by deploying specialized agents designed to assist drivers with personalized, context-aware services. These agents respond to natural language requests and translate them into tractable optimization tasks, enabling real-time adaptation to evolving needs. For instance, a driver might request: “Make sure I have enough charge to attend a late-night event and still commute to work tomorrow”, prompting the system to coordinate a charging plan that balances time constraints, battery wear, and grid conditions. Leveraging the reasoning and coordination capabilities of large language models (LLMs), these agents handle a variety of tasks—including scheduling, prioritization, and decision support—tailored to both individual user preferences and broader system objectives. This design fosters a more natural and effective human-machine interaction model, allowing the IoEV ecosystem to shift away from rigid scheduling schemes toward adaptive and intuitive assistance.



\quad This work proposes a novel Agentic IoEV Intelligence framework that leverages Agentic AI with specialized agents and dynamic analytics tools to address cybersecurity and battery management challenges in the IoEV. Our key contributions are:

\begin{itemize}
\item We propose an Agentic AI architecture for IoEV, centered on a shared, explainable reasoning layer that enables coordination among autonomous agents, supporting secure, reliable, and interpretable electric mobility operations.

\item We develop explainable threat mitigation mechanisms to detect and respond to cyberattacks on charging stations, enhancing IoEV security. Additionally, we propose robust SoC estimation and SoH anomaly detection models with interpretable outputs, supporting proactive maintenance and system reliability. Finally, we design a three-agent pipeline for user-centric support, where each agent leverages LLM-driven reasoning and dynamic tool invocation to interpret intent, contextualize tasks, and execute formal optimizations for user-centric assistance.

\item We conduct comprehensive experiments to evaluate and validate the framework across diverse IoEV scenarios, providing insights into Agentic AI performance. 

\end{itemize}


\quad The rest of the paper is organized as follows: Section~\ref{sec:related} reviews related work on AI for IoEV. Section~\ref{sec:solution} outlines the research questions and proposed Agentic AI framework. Section~\ref{sec:evaluation} details the experimental evaluation and results. Finally, Section~\ref{sec:conclusion} concludes the paper and discusses future directions.

\section{Background \& Related Work}\label{sec:related}

\quad This section reviews key developments across four domains relevant to Intelligent and Secure IoEV: SoC estimation, SoH and anomaly detection, cybersecurity, and agentic AI.

\subsection{SoC Estimation}

\quad Modern SoC estimation combines classical filtering with data-driven models.  Model-based Kalman filters (KF), including extended and unscented variants (EKF/UKF), remain widely used for real-time SoC tracking using physics-based battery models~\cite{xie2023state}, but require accurate parameter identification. Pure machine learning (ML) approaches (e.g., deep neural networks (DNNs), long short-term memory (LSTM) and gated recurrent unit (GRU) networks)~\cite{mousaei2024advancing} are also explored to learn nonlinear battery dynamics from data. Recently, hybrid KF–neural schemes have shown very high accuracy. For example, Liu \& Dun~\cite{liu2024integrated} propose a Dynamic Genetic Kalman Neural Network (DGKNN) that integrates an EKF with a neural net (optimized by a genetic algorithm); this hybrid achieved ~0.15\% SoC prediction error on test data. These hybrids leverage the theoretical reliability of Kalman filtering and the flexibility of learning models to improve real-time SoC monitoring. 




\subsection{SoH Prediction}
\quad SoH prediction (tracking battery aging and capacity) and fault detection have likewise advanced through ML applied to large datasets. A recent study analyzes real-world EV data from 300 vehicles over three years and demonstrates that a deep, multi-modal neural framework can accurately predict battery SoH~\cite{liu2025multi}. In practice, ML-based anomaly detection is used to flag unusual degradation or fault patterns. For example, Cao et al.~\cite{cao2025model} develop a deep-learning (DL)-based battery fault diagnosis network that leverages data from 515 EVs. This model significantly improved the true-positive detection rate of safety-critical faults (e.g., electrolyte leaks, internal shorts) by 46.5\% over previous methods. In summary, data-driven SoH estimators and anomaly detectors use EV operational signals (e.g., voltage, current, temperature) in neural models to identify irregular aging or sensor behavior, yielding both theoretical advances and validated real-world results~\cite{liu2025multi, cao2025model}. DL–based SoH estimation frameworks—often using convolutional neural networks (CNNs) or recurrent neural networks (RNNs)—can learn to predict remaining capacity or degradation trends. For instance, Liu et al.~\cite{liu2025multi} utilize historical vehicle data in a multi-modal DL architecture to achieve efficient and accurate SoH estimation.


\subsection{Cyber Threat Detection} \quad ML techniques are increasingly used for intrusion and anomaly detection in EV-related networks. Two emerging domains are EV charging infrastructure—commonly referred to as electric vehicle charging stations (EVCS)—and in-vehicle networks such as the controller area network (CAN) bus.

\begin{itemize}
\item \textit{EVCS Intrusion Detection Systems (IDS):} Charging stations form an Internet of Things (IoT) ecosystem with standardized communication protocols (e.g., Open Charge Point Protocol (OCPP)) and are vulnerable to cyber-physical attacks. Researchers have applied DNNs with transfer learning (TL) to this problem. For example, Khan et al.~\cite{almadhor2025transfer} train a DNN on a large EVCS attack dataset and then fine-tune it for specific sites. This TL-based detector achieved 98\% accuracy on a public EVCS intrusion dataset. In effect, the model leverages pretrained weights to quickly adapt to new EVCS environments and detect diverse attacks. This work shows that DL-based IDS can significantly outperform traditional signature-based IDS for charging stations by capturing complex and evolving threat patterns.

\item \textit{In-Vehicle (CAN Bus) IDS:} The vehicle’s CAN bus lacks built-in security, making it a key target. Rai et al.~\cite{rai2025securing} demonstrate that modern DL architectures (LSTM, GRU, CNN) can effectively detect CAN-bus intrusions. Using public CAN attack datasets, they report ~99–100\% accuracy in detecting both denial-of-service and spoofing attacks. This study highlights that goal-driven neural agents can analyze the spatio-temporal structure of CAN messages (via bi-directional LSTM (BiLSTM) and CNN models) to catch anomalies that legacy methods miss.
\end{itemize}

\subsection{Agentic AI \& IoEV}
\quad AAI refers to autonomous, goal-directed AI agents that can plan, adapt, and act with minimal human intervention. In robotics~\cite{brohan2023rt}, mission-critical applications supported by 5G/6G networks~\cite{khowaja2025integration}, and smart grids~\cite{acharya2025agentic}, recent work has leveraged such agents (often via reinforcement learning, multi-agent systems, etc.) to achieve autonomous control.


\quad Agentic AI could enhance both battery management and cybersecurity in IoEVs. For instance, an autonomous agent could continuously adapt its SoC/SoH estimation or charging strategy based on driving patterns and grid conditions, effectively optimizing battery life without manual retuning. Likewise, goal-driven agents could monitor vehicle networks and detect intrusions in real time, or even enact countermeasures. Although IoEV-specific agentic systems are still emerging, the successes in robotics (multi-robot with safety guarantees) and energy management suggest strong potential. In short, integrating agentic autonomy into EV systems could enable vehicles to self-manage health and security in a dynamic environment, building on the DL-based IDS and battery management system (BMS) advances noted above.

\section{Methodology}  \label{sec:solution}

\begin{figure*}[t]
    \centering
  \includegraphics[width=0.8\linewidth]{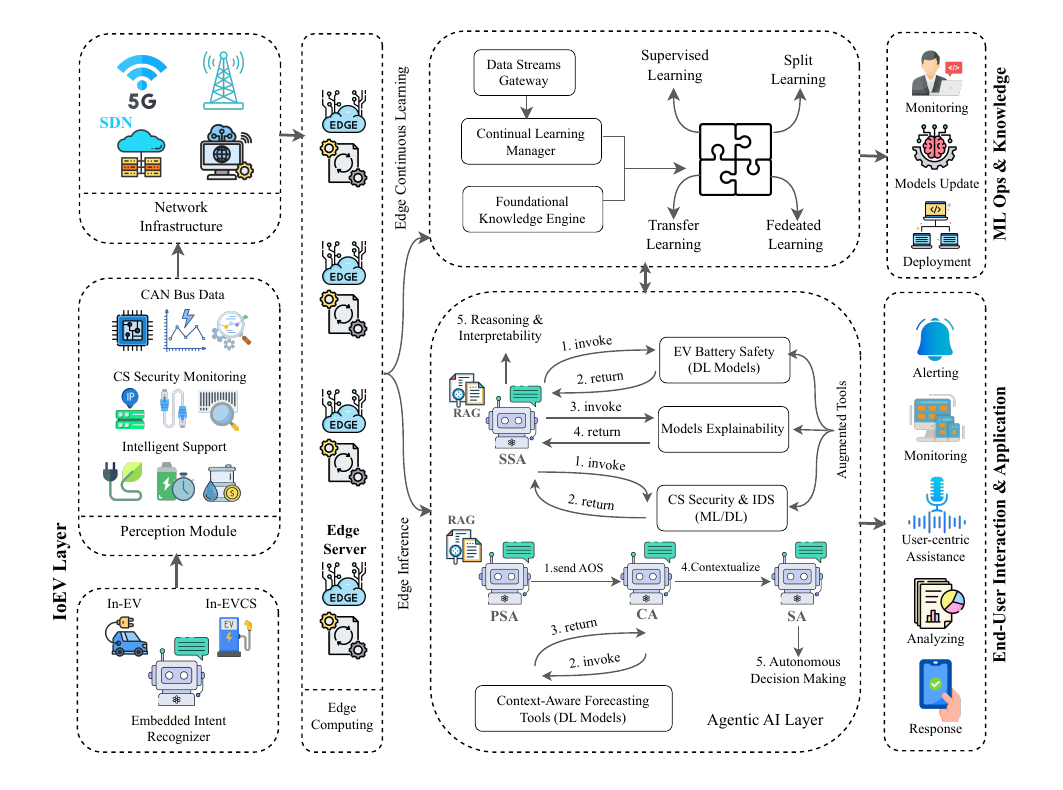}
    \caption{System Architecture — The Agentic IoEV framework adopts a five-layered design comprising the IoEV Layer (data collection from vehicles and charging stations), Network Layer (5G and SDN connectivity), Edge Computing Layer (distributed inference and continual learning), Agentic AI Layer (task-specific intelligent agents), and Application Layer (user interaction and ML Ops).}
    \label{fig:archi}
\end{figure*}


\quad This section outlines the proposed methodology for intelligent and autonomous IoEV systems. We begin by defining the core problem: jointly estimating battery SoC and SoH while detecting cyber threats and supporting end users under real-world uncertainties. To address this, we introduce a five-layer system architecture that supports distributed sensing, computation and decision-making. We then detail the agentic workflow, wherein autonomous agents continuously monitor battery conditions, detect faults and intrusions, and adaptively support end users.

\subsection{Problem Statement}

\quad The increasing connectivity of IoEV introduces significant cybersecurity risks (e.g., DDoS attacks on charging stations) and challenges in reliable battery management (e.g., accurate SoC estimation and SoH anomaly detection). A critical question arises: can Agentic AI with specialized agents and dynamic analytics provide autonomous, explainable, and robust solutions for these challenges? This paper investigates this problem through the following research questions:

\begin{itemize}
\item RQ1: Can Agentic AI accurately detect and mitigate cyberattacks on IoEV charging stations with explainable mechanisms?
\item RQ2: Can Agentic AI provide robust and interpretable SoC/SoH estimations prioritizing critical degradation patterns with interpretable outputs?
\item RQ3: How does the integration of specialized agents and dynamic analytics enhance the adaptability and performance of Agentic AI in dynamic IoEV environments?
\end{itemize}

\subsection{Layered Architecture}


\quad The Agentic IoEV framework combines specialized agents, real-time analytics, and explainable AI within a five-layered architecture that leverages edge computing and 5G connectivity for secure and intelligent decision-making. The design depicted in Figure~\ref{fig:archi} ensures modularity, scalability, and seamless integration of agents, data analytics pipelines, and explainability tools in the IoEV environment. The core functional roles of the layers are outlined below:

\begin{itemize}
    \item IoEV Layer: Aggregates vehicle and charging station data (e.g., voltage, temperature, traffic) to support battery diagnostics and cyber threat detection.
    \item Network Layer: Utilizes 5G and Software-Defined Networking (SDN)\cite{lcnsdn} to ensure low-latency, high-reliability communication for real-time monitoring and response.
    \item Edge Computing Layer: Hosts autonomous agents and supports distributed, low-latency inference~\cite{lcnslicing}, along with continual learning approaches—such as federated, transfer, and split learning—that enable models to adapt incrementally and privately to new data while retaining previously learned knowledge.
    \item Agentic AI Layer: Deploys task-specific agents for cybersecurity, SoC/SoH analytics, and user-centric services using context-aware, explainable reasoning.
    \item Application Layer: Interfaces with users and operators, providing transparency, feedback, and adaptive control through real-time insights, decision oversight and machine learning operations (ML Ops) integration.
\end{itemize}

\subsection{Privacy and Secure Data Handling in Agentic IoEV}

\quad  The Agentic AI framework processes sensitive data—including CAN bus telemetry, calendar-based intent inputs, and charging station traffic—for personalized optimization and threat detection. To protect user privacy and ensure secure edge-based inference, we implement the following safeguards:
\begin{itemize}
    \item \textbf{Secure Transmission of CAN and Network Data:}All telemetry is encrypted using Transport Layer Security (TLS) with mutual authentication and ephemeral session keys to prevent replay or correlation attacks.
    \item \textbf{Minimal Data Retention Policy:}
    Raw data is processed in-session and discarded unless user consent is given for storage.

    \item \textbf{Agent Isolation and Data Access Scoping:}
Agents operate under strict task-specific scopes (e.g., the Contextualizer Agent accesses calendar data, while the Safety Agent handles logs), with fusion only under explicit coordination.
    \item \textbf{Federated Learning with Differential Privacy:}
    To improve task-specific ML models (e.g., SoH predictors, threat classifiers) while safeguarding raw user data, we employ federated learning. Agents contribute encrypted local updates; differentially private noise is applied during aggregation to protect against inference or model inversion attacks.
\end{itemize}
These mechanisms collectively ensure that sensitive data—ranging from vehicle diagnostics and user schedules to charging station logs—is processed securely at the edge, without unnecessary retention, or exposure beyond the agent's authorized scope.
\subsection{In-depth Agentic workflow} 
\quad The Agentic AI Layer operates as a distributed, tool-augmented, and explainable multi-agent system. Each agent within this layer is tailored to handle a specific class of tasks, interfacing with specialized tools, continuously updated models, and explainability components to deliver actionable, human-understandable decisions. The end-to-end workflow begins at user interface level—whether in the EV or the CS—and progresses through intelligent modules designed for perception, communication, and decision-making.

\subsubsection{Embedded Intent Recognizer and Perception} At the EV or charging station level, an Embedded Intent Recognizer—implemented as a compact language model suitable for deployment in resource-constrained environments, such as microcontrollers within EVs monitoring the CAN bus or local control hardware in charging stations—interprets user queries and system events by identifying their underlying intent. This may include a driver's request for battery diagnostics or a system-initiated alert related to anomalies in the charging stations. Upon recognizing the intent, the Perception Module gathers real-time sensor data relevant to the corresponding task and transmits it to the edge infrastructure hosting the Agentic AI components via a high-throughput network.

\subsubsection{Task-Oriented Agentic Routing} Following the perception phase, the system routes data into two main pathways depending on the recognized intent:
\begin{itemize}[left=0pt]
    \item \textbf{Safety and Security Agent (SSA):}
    The SSA is a domain-specialized LLM augmented with Retrieval-Augmented Generation (RAG)~\cite{zhao2024retrieval}, which contains relevant domain-specific documents, and integrated tools for real-time diagnostics, threat analysis, and explainability mechanisms. Upon identifying a safety or security-related request, the SSA acts as a reasoning layer that interprets intent, coordinates tools, and delivers clear, context-aware responses to end users in the following manner:
    \begin{enumerate}[left=0pt, label=\alph*.]
        \item \textbf{Invocation of Diagnostic Tools:} The agent dynamically selects and invokes the appropriate tool based on the recognized intent: \textit{(i) EV Battery Safety Tool,} which is a multi-task pre-trained model designed for real-time inference of the SoH and SoC of EV batteries; \textit{(ii) Charging Station Security Tool,} which utilizes pre-trained anomaly and intrusion detection models to identify potential security breaches or malicious activities.
       
    \item \textbf{Explainability Tool:} After performing diagnostics or threat assessments, the SSA engages an explainability tool—such as SHapley Additive exPlanations (SHAP) or Local Interpretable Model-agnostic Explanations (LIME)—to interpret the results and highlight the key features that most influenced the output. This reveals the underlying factors contributing to degradation or security threats, enabling causal interpretability and providing transparent reasoning about abnormal system behaviors.
    \item \textbf{Response Generation:} The SSA consolidates the results into a user-friendly explanation, ensuring that technical assessments are conveyed in an accessible manner for vehicle users or charging station operators, and providing clear, intuitive responses that are both informative and easily understandable.
    \end{enumerate}
 Through the orchestration of specialized diagnostic tools, explainability techniques, and agents' capabilities, the SSA provides accurate, interpretable, and user-friendly responses in critical safety and security scenarios.

    \item \textbf{Multi-Agent Coordination and Support:}
    When a user issues a high-level, intent-rich request, the system initiates a three-agent pipeline designed to interpret, contextualize, and solve the task through a combination of LLM-driven reasoning and formal optimization methods. This route consists of the Personalized Support Agent (PSA), the Contextualizer Agent (CA), and the Solver Agent (SA). The roles and functions of these agents are as follows:
    \begin{itemize}[left=0pt]
        \item \textbf{Personalized Support Agent:} Responsible for translating a user's natural language request into a formalized optimization problem skeleton and identifying the specific parameters required to instantiate it. The PSA’s first task is to determine the underlying optimization problem type based on the user’s intent. This is accomplished by leveraging the generalization capabilities of an LLM, which maps a free-form user request to one of several predefined optimization problem types. For instance, the request ``\textit{I want to charge my EV tomorrow on my way back home as cheaply as possible}'' would be interpreted as a cost-minimization problem under spatiotemporal constraints, matching a template such as a multi-objective constrained optimization.
Once the problem type is identified, the PSA proceeds to the second task: extracting and categorizing the parameters needed to instantiate the problem. These parameters define the optimization variables, constraints, and objectives, and are organized into two categories:\\
     \textbf{Type 1} - Explicitly stated parameters: Elements directly mentioned in the request, such as ``\textit{tomorrow}'' (temporal constraint) or ``cheap'' (cost minimization objective).\\
    \textbf{Type 2} - Physical system parameters: Inherent properties of the user’s environment or devices necessary for optimization but not typically restated in each request — for example, the electric vehicle’s battery capacity, charging rate, or energy consumption profile.
The PSA outputs the abstract problem skeleton—a symbolic representation of the optimization structure—along with a list of unresolved parameters and their categories. While this representation is not directly solvable, it provides a structured starting point for subsequent contextual grounding and numerical resolution.

\item \textbf{Contextualizer Agent:} Responsible for bridging the gap between symbolic user intent and grounded, data-driven values. It receives the abstract optimization skeleton (AOS) produced by the PSA and enriches it by mapping unresolved symbolic parameters to contextually relevant values. The CA is implemented as an LLM augmented with pretrained deep learning models specialized in forecasting electricity prices, estimating charging station demand and occupancy, and predicting user behavior patterns such as likely return times. These predictions are inferred from user-provided calendar data and historical mobility patterns. To preserve privacy, the CA accesses this data only during active sessions, processes it transiently at the edge, and discards or anonymizes it immediately after use—unless explicit, user-granted retention is required for personalization. Access is strictly scoped, and all identifiers are pseudonymized prior to analysis. This ensures compliance with the system’s minimal data retention and data scoping policies. The CA grounds the abstract formulation with these context-sensitive parameters, producing a fully instantiated optimization problem ready for computational resolution.
\item \textbf{Solver Agent:} Responsible for solving the fully grounded optimization problem delivered by the CA. It uses an LLM combined with a set of external numerical solvers, each specialized for a specific type of optimization problem. Instead of relying only on LLM-generated solutions—which can be approximate and may violate important constraints—the SA uses well-tested solvers for accurate results. Since the PSA has already identified the problem type (such as linear programming, quadratic programming, or mixed-integer optimization), the SA directly selects the appropriate solver from its collection. This collection includes tools from libraries like \texttt{scipy.optimize}, \texttt{cvxpy}, and \texttt{control}. Using these proven algorithms ensures that strict physical and quality-of-service (QoS) constraints are respected.


    \end{itemize}

\end{itemize}

\section{Evaluation \& Results}
\label{sec:evaluation}
\quad This section presents the experiments we conducted to validate the proposed AAI framework for IoEV. 






\subsection{Experimental Setup}
\subsubsection{Datasets and Models} We evaluated several ML/DL models on three distinct tasks using real-world datasets. The first task addressed multitask learning, involving the classification of SoH and the estimation of SoC using the EVBattery dataset \cite{he2022evbattery}. This dataset comprises large-scale time-series charging data, segmented into 128-step sequences using a sliding window, with each snippet containing features such as cell voltage statistics, charging current, temperature, SoC values, and timestamps. The second task focused on detecting cyberattacks using the CICEVSE2024 dataset\cite{buedi2024enhancing}, which captures network traffic from EVCS under \texttt{benign} conditions as well as during \texttt{reconnaissance} and \texttt{denial-of-service} attacks. The raw pcap files were processed using NFStream. To mitigate potential bias, features related to timestamps, Media Access Control (MAC) addresses and Internet Protocol (IP) addresses were removed, and redundant attributes were filtered out based on correlation analysis. The third task involved forecasting EV charging demand in urban environments using the UrbanEV dataset \cite{li2025urbanev}. This dataset includes charging data such as occupancy, duration, and volume, as well as environmental context like weather conditions. Additionally, it incorporates spatial features like adjacency matrices and distances, along with static attributes such as Points of Interest, area size, and road length. Details of the models used and their respective architectures are given in Table~\ref{tab:hyperparams}.


\subsubsection{Evaluation Metrics}  We assess the performance of SoH prediction and EVCS attack detection using accuracy, precision, recall, and F1-score: 

\begin{itemize}

\item[] $\text{Accuracy} = \frac{TP + TN}{TP + TN + FP + FN}$
 \vspace{4pt}
\item[] $\text{Precision} = \frac{TP}{TP + FP} \:\: \textbf{and} \:\: \text{Recall} = \frac{TP}{TP + FN}$
 \vspace{4pt}
\item[] $\text{F1-Score} = 2 \times \frac{\text{Precision} \times \text{Recall}}{\text{Precision} + \text{Recall}}$
 \vspace{4pt}
\end{itemize}

For the SoC estimation and EV/EVCS forecasting tasks, we use the mean absolute error (MAE), the mean squared error (MSE), and the root mean squared error (RMSE):  

\begin{itemize}
\item[]  $ \text{MAE} = \frac{1}{N} \sum_{i=1}^{N} | y_i - \hat{y}_i | $
 \vspace{4pt}
\item[] $ \text{MSE} = \frac{1}{N} \sum_{i=1}^{N} (y_i - \hat{y}_i)^2 \: \: \text{and} \:\:  \text{RMSE} = \sqrt{\text{MSE}}$
\end{itemize}

\subsubsection{Training Configuration}
We perform experiments on three NVIDIA H100 GPUs, repeating each experiment at least 3 times and averaging the results. The default settings for training our models are given in the Appendix~\ref{app:appenmeryem} (Table~\ref{tab:hyperparams}). The agents are implemented in Python using the LangChain library. For the base model, we use LLaMA2-7B, LLaMA2-13B, and LLaMA-8B, as they are open-source and can be run locally with relatively affordable computational resources. To eliminate randomness, the model temperature is set to 0.


\subsection{Intent Recognition}
\begin{figure*}[h]
    \centering
    \begin{subfigure}[b]{0.3\textwidth}
        \includegraphics[width=\textwidth]{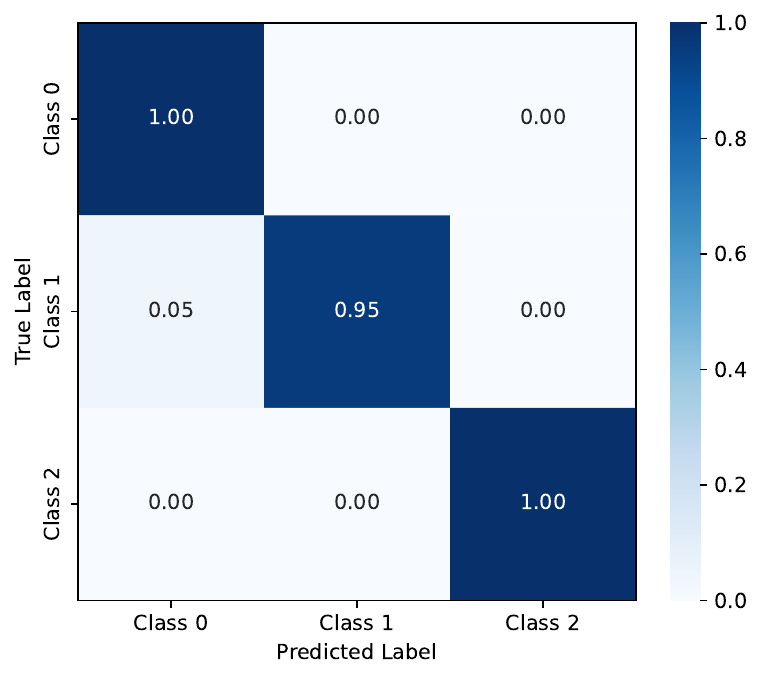}
        \caption{Fine-tuned DistilBERT}
        \label{fig:fig1}
    \end{subfigure}
    \hspace{0.03\textwidth}
    \begin{subfigure}[b]{0.3\textwidth}
        \includegraphics[width=\textwidth]{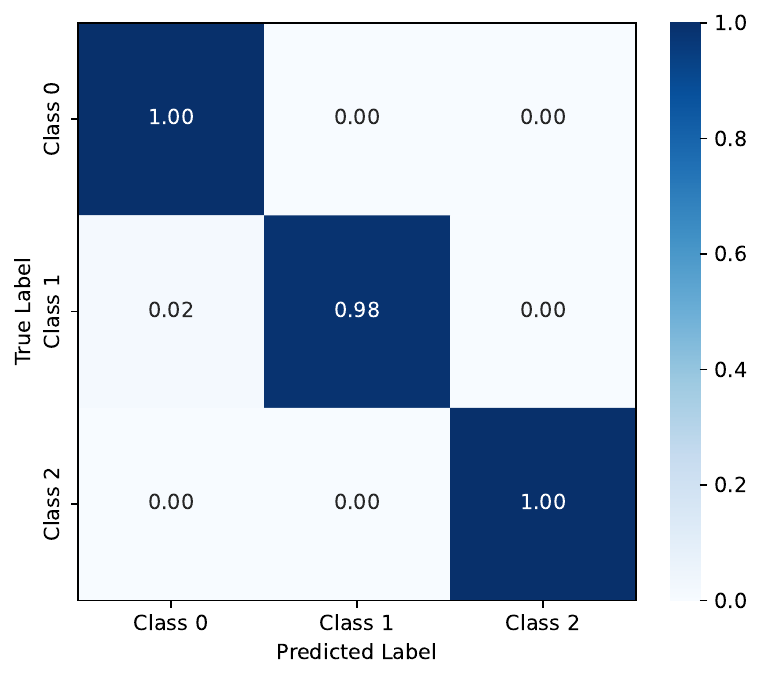}
        \caption{DeBERTa}
        \label{fig:fig2}
    \end{subfigure}
    \hspace{0.03\textwidth}
    \begin{subfigure}[b]{0.3\textwidth}
        \includegraphics[width=\textwidth]{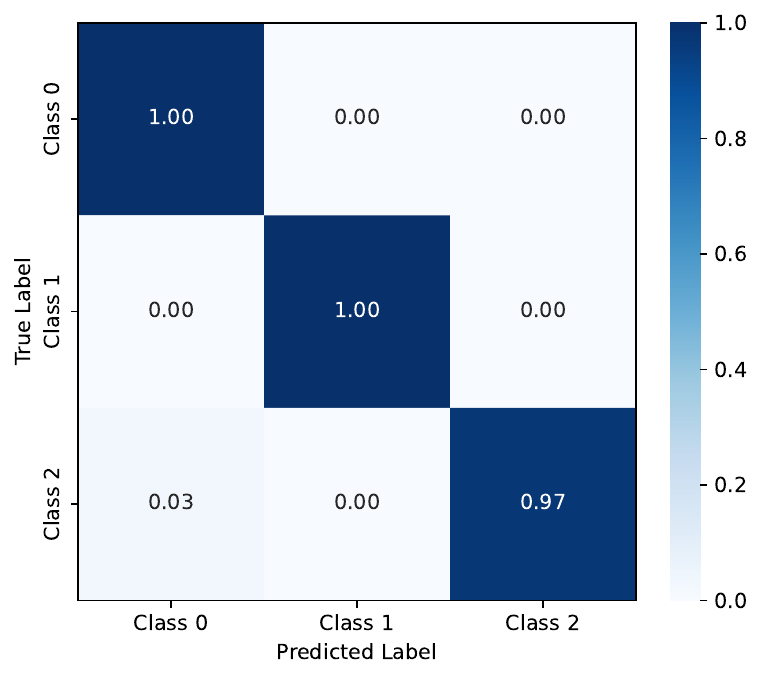}
    \caption{RoBERTa}
        \label{fig:fig3}
    \end{subfigure}
    \caption{Confusion Matrices for Intent Recognition}
    \label{fig:three_figures}
\end{figure*}

We benchmarked our Embedded Intent Recognizer using a synthetic dataset consisting of EV battery diagnostics (class 2), charging station intrusion identification (class 1), and user-centric queries (class 0). A lightweight, fine-tuned DistilBERT achieved performance comparable to larger models like DeBERTa and RoBERTa (as illustrated in the confusion matrices in Figure \ref{fig:three_figures}), excelling at text classification and offering faster inference with significantly lower computational overhead. This makes DistilBERT an ideal choice for resource-constrained environments such as EVs and CSs.

\subsection{Safety and Security}

\begin{table}[th]
\centering
\caption{Performance Comparison of Models for EV Battery State Diagnosis and EVCS Attack Detection: IT: Inference Time and MS: Model Size.}
\label{tab:model_comparison}
\setlength{\tabcolsep}{4pt}
\begin{tabular}{lllccc}
\hline
\textbf{Task} & \textbf{Subtask} & \textbf{Metric} & \textbf{Model 1} & \textbf{Model 2} & \textbf{Model 3} \\
\hline

\multirow{9}{*}{\shortstack[l]{Battery\\State\\Diagnosis}} 
 & \multirow{6}{*}{\shortstack[c]{SoH\\(Anomaly)\\Prediction}} 
   & MS (KB)  & 84.74 & 95.34 & 76.23 \\
   & & IT (Ms)   & 0.2885 & 0.324 & 0.2595 \\
   &  & Accuracy   & 96.12\% & 96.62\% & 94.2\% \\
 & & Recall     & 93.38\% & 93.95\% & 91.30\% \\
 & & F1-score   & 94.69\% & 95.37\% & 92.2\% \\
 \cline{2-6}
 & \multirow{3}{*}{\shortstack[c]{SoC\\(Capacity)\\Estimation}} 
   & MAE        & 0.4625   & 0.4518   & 0.4754   \\
 & & MSE        & 0.6694  & 0.6462  & 0.6706  \\
 & & RMSE       & 0.8182   & 0.8039   & 0.8189   \\
\hline

\multirow{4}{*}{\shortstack[c]{EVCS \\Attacks\\Detection}} 
 & \multirow{4}{*}{\shortstack[c]{\:\:\:\:\:\:---}} 
 & MS(KB)  & 45.4 & 209.5 & 215 \\
   & & IT (Ms)  & 3.017 & 18.15 & 0.214 \\
  & & Accuracy   & 98.12\% & 99.3\% & 98.62\% \\
 & & F1-score   & 98\% & 99.33\% & 98.23\% \\
\hline
\end{tabular}
\label{tab:ml}
\end{table}

\subsubsection{ML/DL Models Performance}
\begin{figure}[th]
    \centering
\includegraphics[width=0.93\linewidth]{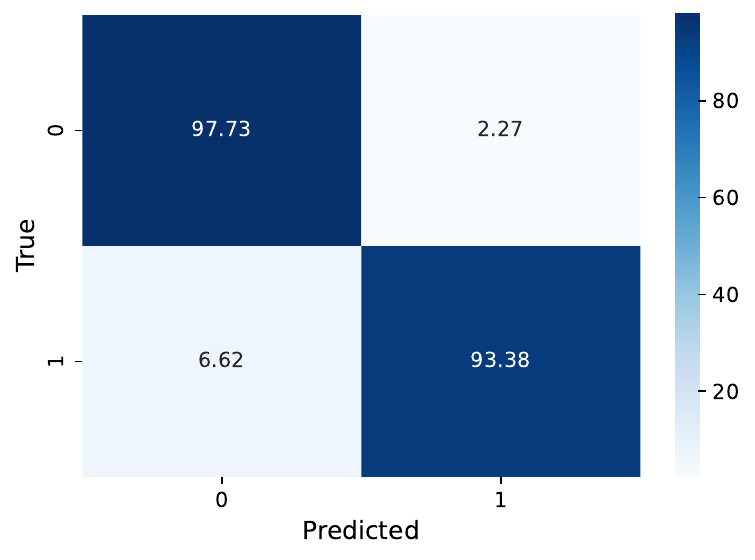} 
    \caption{Confusion Matrix for Binary Classification of EV Battery State of Health}
    \label{fig:appendix-evbcm}
\end{figure}

\begin{figure}[th]
    \centering
\includegraphics[width=0.93\linewidth]{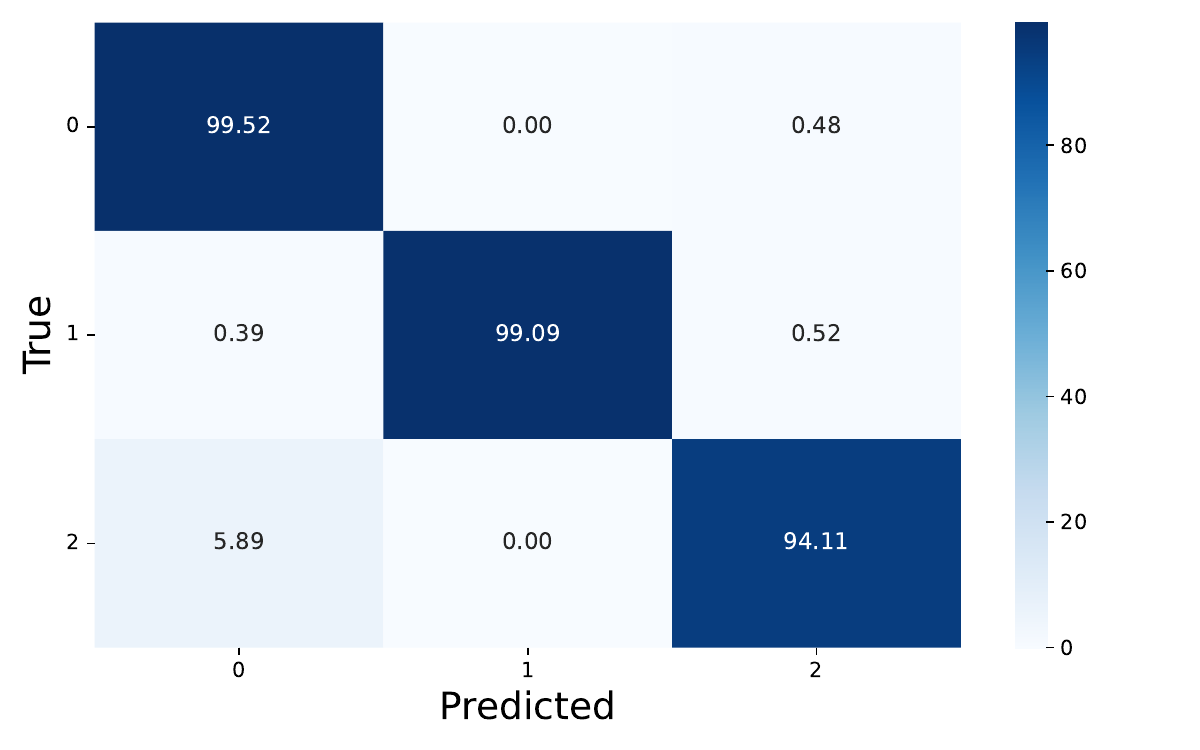} 
    \caption{Confusion Matrix of Predicted vs. Actual Attack Classes on EVCS}
    \label{fig:appendix-evcscm}
\end{figure}

All models were trained and evaluated on the preprocessed datasets, and their performance metrics for both tasks are summarized in Table \ref{tab:ml}. For battery state diagnosis, Model 1 (LSTM), Model 2 (BiLSTM), and Model 3 (GRU) all achieve high performance with low inference time and compact sizes. Similarly, for EVCS attack detection, Model 1 (multi-layer perceptron, MLP), Model 2 (LSTM), and Model 3 (extreme gradient boosting, XGBoost) deliver strong results across all metrics. Evaluating three distinct models per task enables a robust comparison of learning architectures in terms of accuracy, efficiency, and deployment feasibility. Confusion matrices for both classification tasks are provided in Appendix \ref{app:appenmeryem} for additional insight into class-level performance (Figure \ref{fig:appendix-evbcm} for battery state of health diagnosis and Figure \ref{fig:appendix-evcscm} for EVCS attack detection).

\subsubsection{Explainability and Human-Centered Interpretation}
\begin{figure}[th]
    \centering
\includegraphics[width=0.95\linewidth]{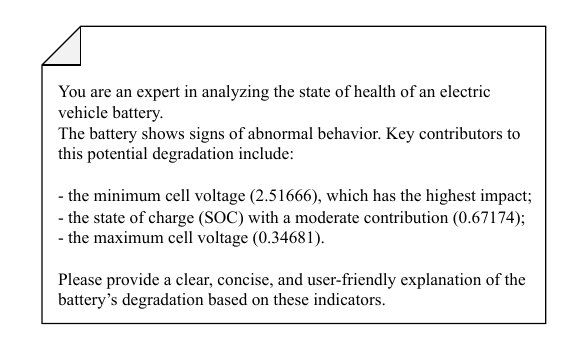} 
    \caption{Prompt Used to Interpret EV Battery Degradation Indicators}
    \label{fig:appendix-evpmt}
\end{figure}

 To enhance transparency and provide interpretable insights for end-users (such as EV drivers and EVCS operators), we incorporated explainability techniques into the outputs of trained ML models using SHAP\cite{mosca2022shap}, a robust tool that calculates the contribution of each feature to the model's prediction, offering a clear understanding of how individual inputs influence the outcome (see Figure \ref{fig:appendix-shap} in the Appendix \ref{app:appenmeryem} for the SHAP summary plot highlighting the features contributing to the classification of an EVCS DoS attack). For each output—whether identifying operational factors leading to EV battery anomalies or detecting security intrusions in the EVCS network—an LLM, enriched with SHAP-derived information, was carefully prompted to generate role-specific explanations (See the prompt for EV battery anomaly interpretability in Figure \ref {fig:appendix-evpmt} and the prompts for EVCS attacks interpretability in Figure \ref{fig:appendix-evcspmt} in the appendix \ref{app:appenmeryem}), ranging from alert messages for drivers regarding battery degradation to incident summaries for EVCS operators. By reasoning through the SHAP outputs, the LLM provided human-understandable explanations tailored to the needs of different stakeholders, ensuring clarity and actionable insights while minimizing reliance on expert or domain-specific knowledge and intervention.
 
 \subsubsection {SSAs Explainability Performance}

\begin{table}[th]
\centering
\caption{Accuracy-based BARTScore Comparison of SSA's Responses using LLaMA Models}
\setlength{\tabcolsep}{4pt}
\begin{tabular}{lccc}
\toprule
\textbf{Task} & \textbf{LLaMA 7B} & \textbf{LLaMA 13B} & \textbf{LLaMA3 8B} \\
\midrule
EV Battery Checkup   &  \textbf{0.84}     &  \textbf{0.87} & \textbf{0.95} \\
EVCS Attacks           &  \textbf{0.86}     &  \textbf{0.89} & \textbf{0.93} \\
\bottomrule
\end{tabular}
\label{tab:bartscore}
\end{table}

 To assess the interpretative capability of the LLM used as SSA, we conducted a two-part evaluation corresponding to each task. From the EVBattery\cite{he2022evbattery} and CICEVSE2024\cite{buedi2024enhancing} datasets, we extracted 100 random samples each, processed them through the trained ML models, and applied SHAP to highlight the most influential input features. These SHAP-based insights were then transformed into natural language prompts designed for GPT-4, simulating realistic queries from end-users such as EV drivers or EVCS operators. The generated explanations were subsequently annotated and verified by domain experts, who assessed the correctness, clarity, and relevance of each explanation. The evaluation showed that 98\% of the responses were judged to be accurate and meaningful, indicating that the GPT-4 can provide high-fidelity, expert-like interpretations and may be considered as ground truth references for evaluating our deployed models, specifically, prompt-engineered LLaMA2-7B, LLaMA2-13B and LLaMA3-8B models. We prioritized these smaller models not merely due to hardware constraints, but to ensure lower energy consumption, and cost-effective scalability across multiple edge servers.
Each agent was prompted with the same input, and its generated responses were compared with expert-verified GPT-4 references using BARTScore to assess linguistic quality and semantic similarity~\cite{bouchiha2024llmchain}.
Prior to adopting BARTScore for large-scale evaluation, we validated its reliability by comparing its outputs with expert human judgments. Domain experts assessed each response to determine its alignment with the ground truth. Then these expert evaluations were compared with automated BARTScore evaluations, which resulted in a strong correlation of \textbf{0.91}. This confirmed BARTScore's ability to reflect expert evaluations accurately, thereby justifying its use for systematic quality assessment. This approach ensured that the automatic review mirrored the expert assessments, offering confidence that BARTScore can be reliably used for quantifying the quality of LLM-generated responses.
In addition, exact match rates were calculated to quantify accuracy in reproducing the core content of the reference. As illustrated in Table~\ref{tab:bartscore}, the LLaMA 7B, 13B, and 3 8B models demonstrated strong performance on both tasks, achieving accuracies of 0.84, 0.87, and 0.948 for the EV Battery Checkup task, and 0.86, 0.89, and 0.93 for the EVCS Attacks task, respectively. These results highlight that effective prompt engineering, without the need for costly fine-tuning, can lead to high-quality results. This evaluation allowed us to gauge how well smaller and more recent LLMs can produce faithful and informative explanations, using GPT-4 outputs as a reliable expert-level reference.

\subsection{User Support}
\begin{table}[th]
\centering
\caption{Evaluation of PSA Performance}
\setlength{\tabcolsep}{4pt}
\begin{tabular}{lccc}
\toprule
\textbf{Task} & \textbf{LLaMA 7B} & \textbf{LLaMA 13B} & \textbf{LLaMA3 8B} \\
\midrule
AOS Extraction   &  \textbf{0.80}     &  \textbf{0.84} &  \textbf{0.91} \\
\bottomrule
\end{tabular}
\label{tab:psa}
\end{table}

\quad We benchmarked the coordination and support capabilities of our multi-agent system by evaluating the two core agents individually: the PSA and the CA. The focus is placed on these two agents, as the third agent, the SA, merely maps the finalized problem and its instantiated parameters to external python solvers tailored for optimization tasks and does not contribute to the reasoning or abstraction processes that are central to our evaluation.

\quad The PSA translates end-user assistance requests into formal optimization problem types and identifies the parameters required to instantiate them. To assess its performance, we created a synthetic dataset comprising diverse user-centric queries related to charging optimization across a range of real-world scenarios and constraints. These prompts were passed to GPT-4, which generated corresponding optimization problem types along with their skeletons, including parameters and constraints. The outputs were manually reviewed and annotated based on their correctness, relevance, and alignment with the prompt’s intent. Only those deemed accurate and suitable were retained as ground truths, resulting in an accuracy of \textbf{97.2\%} This process demonstrates that GPT-4 can effectively serve as a ground truth generator for this task.
Subsequently, the same prompts were given to prompt-engineered LLaMA-7B, 13B, and 3 8B models, which are lighter-weight alternatives to GPT-4, to evaluate their ability to extract AOS from user intents. The generated outputs were compared to the GPT-4-based ground truths using an automatic evaluation procedure, where we employed DeepSeek\cite{guo2025deepseek} as the judge model. Its role was to assess whether the candidate outputs matched the ground truth in terms of both structure and semantic alignment. We opted to use DeepSeek as the judge model instead of automated scoring metrics such as BARTScore, as the task of extracting an optimization skeleton from user intent requires contextual reasoning and structural understanding, which are beyond the capabilities of token-level similarity measures. The resulting accuracies, reported in Table~\ref{tab:psa}, indicate that while LLaMA-13B outperformed LLaMA-7B, LLaMA 3 8B achieved the best performance with an accuracy of 0.91. These findings suggest that more recent LLMs, even without domain-specific fine-tuning, can effectively approximate expert-level reasoning. However, further improvements may be possible through targeted adaptation, which we leave to future work.

\quad For the CA, which determines whether real-world contextual information is needed and maps abstract optimization parameters to real-life practical, contextually relevant values, we augmented the LLM with external urban EV/EVCS-related tools. These tools include components for forecasting electricity prices, estimating charging station demand and occupancy, and predicting user behavior patterns, such as expected return times inferred from calendar data or mobility habits. The agent invokes these tools only when real-life grounding is required. The forecasting components rely on a pre-trained LSTM model, which estimates charging station demand, occupancy, duration, volume, and electricity prices. For this case study, occupancy, demand, and volume were forecasted based on six time measurements. Table \ref{tab:ca} provides the RMSE and MAE for the forecasting components, which demonstrate good prediction performance, though the error values differ due to the varying scales of the data being forecasted. 
\begin{table}[ht]
\caption{Forecasting model performance for different components}
\centering
\begin{tabular}{lcc}
\hline
\textbf{Forecasting Component} & \textbf{RMSE} & \textbf{MAE} \\
\hline
Charging Station Occupancy & 0.09 & 0.075 \\
Charging Duration & 3.02 & 2.17 \\
Charging Volume & 42.17 & 35.17 \\
\hline
\end{tabular}

\label{tab:ca}
\end{table}





\section{Conclusion and Future Work}
\label{sec:conclusion}
\quad In this work, we presented a novel Agentic AI framework that empowers the Internet of Electric Vehicles with autonomous cyber-threat defense, robust battery-state analytics, and transparent decision support. By decomposing responsibilities across specialized agents—for threat mitigation, SoC estimation, and SoH anomaly detection—and unifying them under an explainable reasoning layer, our approach demonstrably enhances security, prediction accuracy, and stakeholder trust. Extensive evaluations across varied IoEV scenarios confirm substantial improvements over non-agentic methods. Future research will explore dynamic agent collaboration strategies, real-world field trials, and integration with emerging V2G standards to further advance resilient, trustworthy electric-mobility ecosystems. In addition, we plan to investigate the adversarial resilience of LLM-driven agents, focusing on their robustness against realistic threats such as prompt injection, jailbreaking, adversarial intent manipulation, and input poisoning. Strengthening the security posture of autonomous agents will be critical for dependable deployment in safety-critical IoEV applications.

\section*{Acknowledgment}
\quad This work is supported by the OPEVA project, which has received funding within the Chips Joint Undertaking (Chips JU) from the EU’s Horizon Europe Programme and the National Authorities (France, Czechia, Italy, Portugal, Turkey, Switzerland), under grant agreement 101097267. In France, the project is funded by BPI France under the France 2030 program on ``Embedded AI''. Views and opinions expressed are, however, those of the authors only and do not
necessarily reflect those of the EU or Chips JU. Neither the EU nor the granting authority can be held responsible for them.

\bibliographystyle{IEEEtran}
\bibliography{refs}


\section*{Appendix}
\label{app:appenmeryem}
This appendix contains supplementary figures and a table supporting the main analysis.
\begin{figure}[th]
    \centering
\includegraphics[width=1\linewidth]{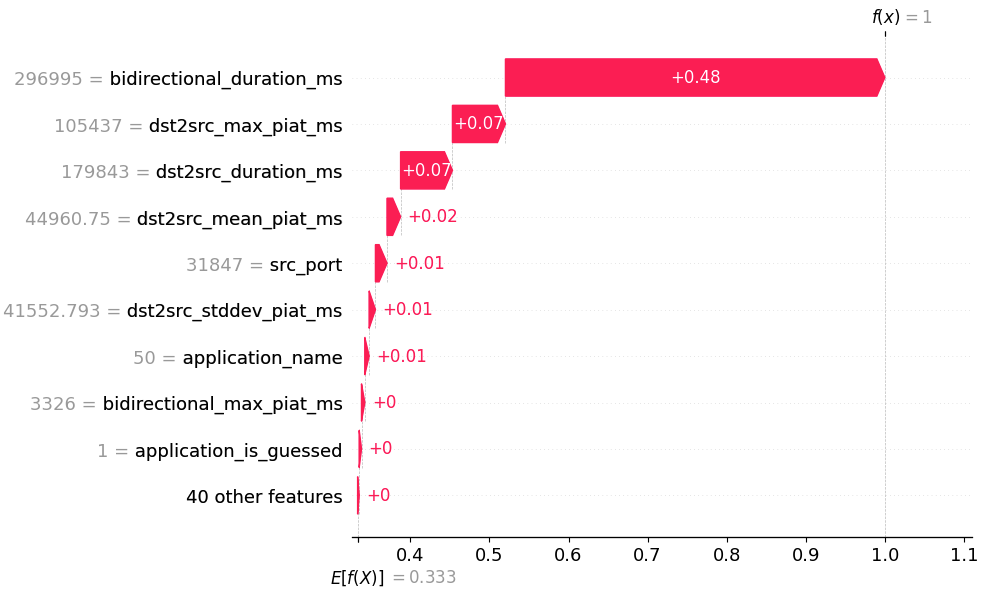} 
    \caption{SHAP Waterfall Plot for a DoS Attack Sample. The plot illustrates the contribution of the top features in pushing the model's output toward classifying the input as a DoS attack. The x-axis represents SHAP values, which quantify each feature’s impact on the model’s final prediction. The y-axis lists the most influential input features, each annotated with the actual feature value from the input sample. The base value E[f(x)] is the model’s average prediction over all training data, while the final output f(x) is the model’s specific prediction for this sample. Features like \texttt{bidirectional\_duration\_ms} (the duration of the bidirectional communication in milliseconds) significantly increase the model's confidence in predicting a DoS attack.}
    \label{fig:appendix-shap}
\end{figure}
\begin{figure}[th]
\centering
\includegraphics[width=0.95\linewidth]{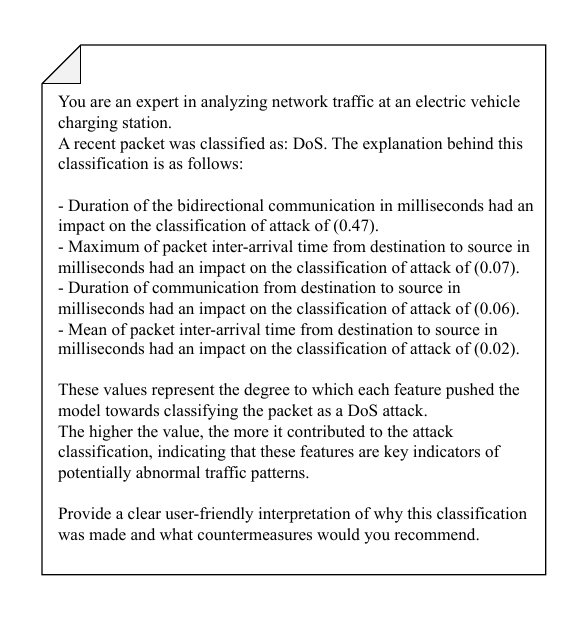} 
    \caption{Interpretability Prompt for EVCS Network Attack Detection}
    \label{fig:appendix-evcspmt}
\end{figure}
\begin{table*}[th]
\centering
\caption{Hyperparameter Configuration for All ML/DL Models}
\begin{tabular}{c c >{\centering\arraybackslash}p{8cm}} 
\toprule
\textbf{Task} & \textbf{Model} & \textbf{Hyperparameters}  \\
\midrule
\multirow{3}{*}{Battery
State
Diagnosis} 
    & MultiTaskLSTM & [Number of layers: 2, Hidden units: 64, Classifier layers: 2 (32, 1), Regressor layers: 2 (32, 1), Dropout: 0.3, Learning rate: 0.001, Batch size: 8, Rounds Number: 20, Proximal regularization: 0.2, Early stopping: True, Patience: 10] \\ \cline{2-3}
    
     & MultiTaskGRU & [Same as MultiTaskLSTM] \\ \cline{2-3}
    & MultiTaskBiLSTM & [Number of layers: 2, Hidden units: 64 (per direction), Classifier layers: 2 (32, 1), Regressor layers: 2 (32, 1), Dropout: 0.3, Learning rate: 0.001, Batch size: 8, Rounds number: 20, Proximal regularization: 0.2, Early stopping: True, Patience: 10] \\
   
\midrule
\multirow{3}{*}{EVCS Attacks
Detection} 
    & MLP & [Number of layers: 4, Neurons per layer: (16, 128, 64, 3)] \\ \cline{2-3}
    & LSTM & [Number of layers: 3, Units per layer: (100, 50, 3)] \\ \cline{2-3}
    & XGBoost & [Number of estimators: 500, Max tree depth: 5, learning rate: 0.07, subsample fraction: 0.65, Column subsample rate: 0.67, Min child weight: 9] \\
\midrule
\multirow{1}{*}{EV-EVCS Forecasting} 
    & LSTM & [Number of layers: 2, Hidden units: 16, Sequence length: (3,6,9,12)]\\
\bottomrule
\end{tabular}
\label{tab:hyperparams}
\end{table*}
\end{document}